\documentclass[12pt]{article}
\usepackage{epsfig,amsmath,amssymb}
\usepackage{color}
\usepackage{graphicx}
\usepackage{cite}
\usepackage{epsfig,amsmath,amssymb}
\usepackage{color}

\begin{document}
\title{Thermodynamic properties of the 2D frustrated Heisenberg model
for the entire $J_{1}-J_{2}$ circle}
\author{{\small A.V. Mikheyenkov$^{1,2,3}$ \thanks{mikheen@bk.ru}, A.V. Shvartsberg$^2$, V.E. Valiulin$^2$ and A.F. Barabanov$^1$}\\
{\small $^1$ Institute for High Pressure Physics RAS, 142190 Moscow (Troitsk), Russia}\\
{\small $^2$ Moscow Institute of Physics and Technology, 141700 Dolgoprudny, Russia}\\
{\small $^3$ National Research Centre 'Kurchatov Institute', 123182 Moscow, Russia}
}
\maketitle
\begin{abstract}
Using the spherically symmetric self-consistent Green's function method, we consider thermodynamic
properties of the $S=1/2$ $J_1$-$J_2$ Heisenberg model on the 2D square lattice. We calculate the
temperature dependence of the spin-spin correlation functions $c_{\mathbf{r}}=\langle
S_{\mathbf{0}}^{z}S_{\mathbf{r}}^{z}\rangle $, the gaps in the spin excitation spectrum, the energy
$E$ and the heat capacity $C_{V}$ for the whole $J_{1}$--$J_{2}$-circle, i.e. for arbitrary
$\varphi$, $J_1=cos(\varphi)$, $J_2=sin(\varphi)$. Due to low dimension there is no long-range
order at $T\neq 0$, but the short-range holds the memory of the parent zero-temperature ordered
phase (antiferromagnetic, stripe or ferromagnetic). $E(\varphi)$ and $C_{V}(\varphi)$ demonstrate
extrema "above" the long-range ordered phases and in the regions of rapid short-range rearranging.
Tracts of $c_{\mathbf{r}}(\varphi)$ lines have several nodes leading to nonmonotonic
$c_{\mathbf{r}}(T)$ dependence. For any fixed $\varphi$ the heat capacity $C_{V}(T)$ always has
maximum, tending to zero at $T\rightarrow 0$, in the narrow vicinity of $\varphi = 155^{\circ}$ it
exhibits an additional frustration-induced low-temperature maximum. We have also found the
nonmonotonic behaviour of the spin gaps at $\varphi=270^{\circ}\pm 0$ and exponentially small
antiferromagnetic gap up to ($T\lesssim 0.5$) for $\varphi \gtrsim 270^{\circ}$.
\end{abstract}

\noindent PACS codes\\
75.10.Jm Quantized spin models, including quantum spin frustration\\
75.10.Kt Quantum spin liquids, valence bond phases and related phenomena\\
75.30.Kz Magnetic phase boundaries

\setcounter{tocdepth}{2}
\tableofcontents

\section{Introduction}
Investigation of low-dimensional quantum magnets have attracted
much attention during the last years (see \cite{Diep13_WS} for a
review). Frustrated two-dimensional (2D) and quasi-2D magnets are
of particular interest, as they demonstrate strong quantum
fluctuations effects. The 2D spin-$1/2$ $J_1$--$J_2$ quantum
Heisenberg model is a conventional tool for the investigation of
frustration effects and quantum phase transitions (see, e.g.,
\cite{Chandr88_PRB,Schulz92_EL,Trumpe97_PRL,Caprio01_PRL,Singh03_PRL,Baraba11_TMP}).

Cuprates and numerous other quasi-2D compound with antiferromagnetic (AFM) nearest-neighbour (NN)
and next-nearest neighbour (NNN) couplings $J_1>0, J_2>0$ have been investigated experimentally for
years \cite{Melzi00_PRL,Melzi01_PRB,Rosner03_PRB,Tranqu07_BookChap}

This class of systems has been recently expanded by several
magnetic materials with a ferromagnetic (FM) NN coupling $J_1<0$
and a frustrating AFM NNN coupling $J_2>0$, e.g.,
Pb$_2$VO(PO$_4$)$_2$
\cite{Kaul04_JMMM,Skoula07_JMMM,Carret09_PRB,Skoula09_EL},
(CuCl)LaNb$_2$O$_7$ \cite{Kageya05_JPSJ}, SrZnVO(PO$_4$)$_2$
\cite{Skoula09_EL,Tsirli09_PRBa,Tsirli09_PRB}, and
BaCdVO(PO$_4$)$_2$ \cite{Carret09_PRB,Tsirli09_PRBa,Nath08_PRB}.
The frustrating $J_2$ is believed to be large enough to drive
these materials out of the FM phase. There are also
materials, such as K$_2$CuF$_4$, Cs$_2$CuF$_4$, Cs$_2$AgF$_4$,
La$_2$BaCuO$_5$, and Rb$_2$CrCl$_4$,
\cite{Feldke95_PRB,Feldke98_PRB,Manaka03_PRB,Kasina06_PRB} with
insufficiently strong frustrating AFM NNN interaction.

\begin{figure}[tbp]
\begin{center}
\includegraphics[width=.6\textwidth]{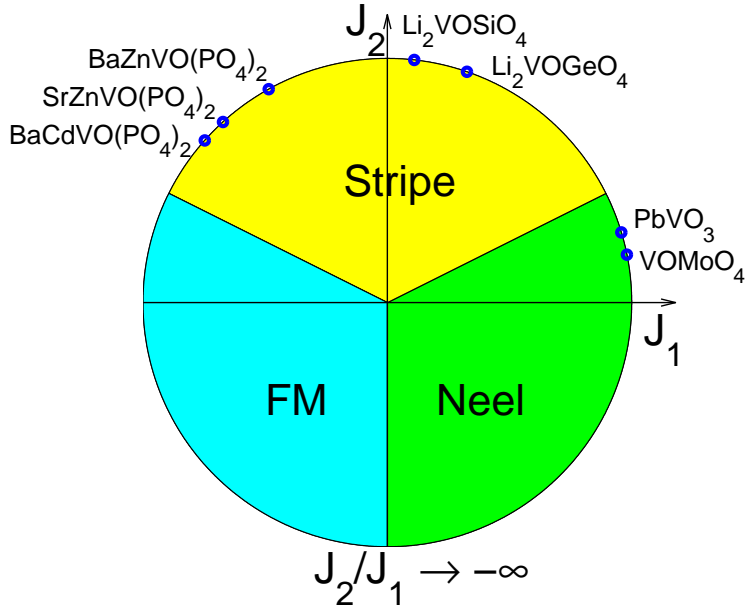}
\caption{(Color online) Circle phase diagram of the 2D J1-J2
Heisenberg model in classical limit; points on the circle
correspond to exchange parameters of several layered compounds,
data from \cite{Nath08_PRB}} \label{fig:fig1}
\end{center}
\end{figure}

The general picture can be seen from Figure~\ref{fig:fig1}, where the phase diagram of 2D
$J_1$--$J_2$ model in the classical limit $S \to \infty$ is complemented by the positions of
several experimental systems. Hereinafter the $J_1$--$J_2$-circle is defined by $J_1=cos(\varphi)$,
$J_2=sin(\varphi)$. In the classical limit only three phases are realized --- all with long-range
order (LRO) --- AFM, FM and stripe (in the quantum case $S=1/2$ disordered phases appear between
stripe and FM, as well as between stripe and AFM).

The Hamiltonian of the model has the form
\begin{equation}
H=J_{1}\sum_{\langle\mathbf{i},\mathbf{j}\rangle}
\widehat{\mathbf{S}}_{\mathbf{i}}\widehat{\mathbf{S}}_{\mathbf{j}}+
J_{2}\sum_{[\mathbf{i},\mathbf{j}]}\widehat{\mathbf{S}}_{\mathbf{i}}
\widehat{\mathbf{S}}_{\mathbf{j}}  \label{Ham}
\end{equation}
where $(\widehat{\mathbf{S}}_{\mathbf{i}})^2=3/4$,
$\langle\mathbf{i},\mathbf{j}\rangle$ denotes NN bonds and
$[\mathbf{i},\mathbf{j}]$ denotes NNN bonds of the square lattice
sites $\mathbf{i},\mathbf{j}$.

The theoretical investigation of the model in the first quadrant
$J_1>0, J_2>0$ was detonated by the HTSC breakthrough and led to
innumerable number of papers (see, e.g. \cite{Richte04_BookChap,
Plakid10_BookChap} and references therein). In a nutshell, the
generally accepted result is the following. At $T=0$ the system
undergoes two successive phase transitions: from AFM LRO
to disordered phase and then to stripe LRO
(see, for example, recent calculations \cite{Ren14_JPCM} and
references therein). The nature of these quantum phase transitions
and the detailed structure of the disordered state remains
debatable.

The unfrustrated FM case ($J_1<0, J_2=0$) has been also
widely investigated, e.g., by the modified spin-wave theory
\cite{Takaha86_PTPS,Takaha87_PRL}, renormalization group
approaches \cite{Kopiet89_PRB,Karche97_PRB}, the quantum Monte
Carlo method
\cite{Juhasz08_PRB,Manous89_PRB,Chen91_PRB,Heneli00_PRB} and by a
spherically symmetric self-consistent approach (SSSA) ---
spin-rotation-invariant Green's function method (RGM) in
alternative notation ---
\cite{Kondo72_PTP,Shimah91_JPSJ,Suzuki94_JPSJ,Junger04_PRB,Antsyg08_PRB}.

Recent experiments with vanadates and related compounds have
stimulated theoretical studies of the $J_1$--$J_2$ model with
$J_1<0$ and frustrating $J_2>0$.
\cite{Shanno04_EPJB,Shanno06_PRL,Sindzi07_JMMM,
Schmid07_JPCM,Schmid07_JMMM,Viana07_PRB,Sindzi09_JPCS,
Shindo09_PRB,Hartel10_PRB,Hartel13_PRB,Dmitri97_PRB,Richte14_JPCS}.
It was found that in the second quadrant there is also a
disordered ground state between FM and stripe. Rough estimates for
the transition points in both quadrants are $J_2 \sim \pm 0.4J_1$
(AFM $\to$ disordered, FM $\to$ disordered) and $J_2 \sim \pm
0.7J_1$ (disordered $\to$ stripe ). Note that for the classical
model ($S \to \infty$) there are only two transitions at points
$J_2=\pm 0.5J_1$ (AFM $\to$ stripe, FM $\to$ stripe).

So, one has several experimental points settled on the upper half
of the $J_1$--$J_2$-circle and a set of theoretical methods, each
being tuned for a particular parameter region. A unified approach,
which can describe the entire picture, both for the ground state
and the thermodynamics of the model, is obviously desirable. It is
also interesting to look at the lower half of the circle
($J_2<0$), though still experimentally unobtainable.

The SSSA proved to be the appropriate approach. SSSA preserves the
spin SU(2) and translation symmetries of the Hamiltonian and
allows:

\textbf{i.} to satisfy automatically the Marshall and
Mermin–Wagner theorems

\textbf{ii.} to describe at $T=0$ (when the LRO is possible) the
system states both with and without LRO in the frames of one and
the same approach

\textbf{iii.} to find the microscopic characteristics such as the
spin-excitation spectrum $\omega(\mathbf{q})$, the spin-gaps
$T$-dependence and the explicit form of the  dynamic
susceptibility $\chi(\mathbf{q},\omega,T)$; to go beyond the
mean-field approximation by introducing damping in the expression
for the spin Green's function $G(\mathbf{q},\omega$)
\cite{Baraba11_TMP}.

Note, that for the 1D spin-$1/2$ Heisenberg ferromagnet it was
shown that the RGM reproduces Bethe-ansatz results
\cite{Suzuki94_JPSJ,Hartel08_PRB}.

Let us also mention that SSSA has been applied to another lattice
geometry, $S>1/2$ \cite{Rubin12_PLAa}, the systems with the
anisotropic spin exchange
--- iron pnictides AFe$_2$As$_2$ (A = Ca, Sr, Ba) ---
\cite{Vladim14_EPJB} and to 2D $J_1$--$J_2$--$J_3$ model
\cite{Baraba11_TMP,Mikhey11_JL,Mikhey12_SSC}.

We have recently considered the ground state of the model
(\ref{Ham}) $J_{1}-J_{2}$ circle \cite{Mikhey13_JL}. In
particular, it was shown that at $T=0$ the transitions between all
ordered and disordered phases are continuous, except the
transition FM$\rightarrow $AFM at $J_{1}=0$, $J_{2}=-1$.

In the present work we turn to the nonzero temperatures and
consider thermodynamic properties. We calculate the temperature
dependence of the spin-spin correlation functions, the gaps in the
spin excitation spectrum and the heat capacity. As in
\cite{Mikhey13_JL}, the model is treated in SSSA.

The paper is organized as follows: in Sec.~\ref{sec:SSSA} and Sec.~\ref{sec:GroSt} we
briefly remind the spherically symmetric self-consistent approach
for two-time retarded Green's functions and the results of
\cite{Mikhey13_JL} for $T=0 $. In Sec.~\ref{sec:Thermod} we present the results
for the thermodynamic properties. Discussion and summary are given
in Sec.~\ref{sec:Diss}.

\section{Spherically symmetric self-consistent approach}
\label{sec:SSSA}

As already mentioned, the calculations are performed in SSSA for the
spin-spin Green's function
\cite{Baraba11_TMP,Kondo72_PTP,Shimah91_JPSJ,
Hartel10_PRB,Hartel13_PRB,Baraba94_JETP,Baraba94_JPSJ}.
\begin{equation}
G_{\mathbf{nm}}=\langle S_{\mathbf{n}}^{z}|S_{\mathbf{m}}^{z}\rangle
_{\omega +i\delta }=-i\int\limits_{0}^{\infty }dt\,e^{i\omega t}\langle
\lbrack S_{\mathbf{n}}^{z}(t),S_{\mathbf{m}}^{z}]\rangle
\end{equation}

Due to the spherical symmetry, only the Green's functions diagonal with
respect to $\alpha =x,y,z$ are nonzero, mean cite spin is zero
($G^{zz}=G^{xx}=G^{yy}$; $\langle S_{\mathbf{n}}^{\beta }\rangle =0,$ $\beta
=x,y,z$). There are three branches of spin excitations degenerate with
respect to $\beta $. Because $\langle \left[
S_{\mathbf{n}}^{z},S_{\mathbf{m}}^{z}\right] \rangle =0$ and $[S_{\mathbf{n}}^{\alpha
};S_{\mathbf{m}}^{\beta }]=i\delta _{\mathbf{nm}}\varepsilon _{\alpha \beta \gamma
}S_{\mathbf{n}}^{\gamma }$, we have for $G_{\mathbf{nm}}$
\begin{equation}
\omega G_{\mathbf{nm}}=\langle \left[
S_{\mathbf{n}}^{z},S_{\mathbf{m}}^{z}\right] \rangle +\langle \lbrack
S_{\mathbf{n}}^{z},\hat{H}]|S_{\mathbf{m}}^{z}\rangle _{_{\omega +i\delta
}}=i\sum_{\mathbf{b}=\mathbf{g},\mathbf{d}}J_{\mathbf{b}}\langle
(S_{\mathbf{n+b}}^{x}S_{\mathbf{n}}^{y}-S_{\mathbf{n+b}}^{y}S_{\mathbf{n}}^{x})
|S_{\mathbf{m}}^{z}\rangle _{_{\omega +i\delta }}
\label{a_Eq1_GF}
\end{equation}
where $J_{g}=J_{1},$\ $J_{d}=J_{2}$. The second differentiation step leads
to three-site Green's functions:
\begin{equation*}
i\omega \sum_{\mathbf{b}=\mathbf{g},\mathbf{d}}J_{\mathbf{b}}\langle (S_{%
\mathbf{n+b}}^{x}S_{\mathbf{n}}^{y}-S_{\mathbf{n+b}}^{y}S_{\mathbf{n}%
}^{x})|S_{\mathbf{m}}^{z}\rangle _{_{\omega +i\delta }}=2\sum_{\mathbf{b}=%
\mathbf{g},\mathbf{d}}J_{\mathbf{b}}C_{\mathbf{b}}(\delta _{\mathbf{n+b},%
\mathbf{m}}-\delta _{\mathbf{n},\mathbf{m}})-
\end{equation*} \begin{equation*}
-\frac{1}{2}\sum_{\mathbf{b}=\mathbf{g},\mathbf{d}}J_{\mathbf{b}}^{2}(G_
{\mathbf{n+b,m}}-G_{\mathbf{nm}})-\sum_{{ \mathbf{b},\mathbf{b}^{\prime };
\beta \neq z;\,\mathbf{b}+\mathbf{b}^{\prime }\neq 0}}
J_{\mathbf{b}}J_{\mathbf{b}^{\prime }}[\langle S_{\mathbf{n+b}
}^{\beta }S_{\mathbf{n}-\mathbf{b}^{\prime }}^{z}S_{\mathbf{n}}^{\beta }|S_
{\mathbf{m}}^{z}\rangle _{_{\omega +i\delta }}-
\end{equation*} \begin{equation}
-\langle S_{\mathbf{n+b}}^{\beta }S_{\mathbf{n}-\mathbf{b}^{\prime }}^{\beta
}S_{\mathbf{n}}^{z}|S_{\mathbf{m}}^{z}\rangle _{_{\omega +i\delta }}+\langle
S_{\mathbf{n+b}}^{z}S_{\mathbf{n+b+b}^{\prime }}^{\beta }S_{\mathbf{n}%
}^{\beta }|S_{\mathbf{m}}^{z}\rangle _{_{\omega +i\delta }}-\langle S_{%
\mathbf{n+b}}^{\beta }S_{\mathbf{n+b+b}^{\prime
}}^{z}S_{\mathbf{n}}^{\beta }|S_{\mathbf{m}}^{z}\rangle _{_{\omega
+i\delta }}]\}  \label{a_Eq2_GF}
\end{equation}

where $c_{\mathbf{b}}=\langle
S_{\mathbf{n}+\mathbf{b}}^{z}S_{\mathbf{n}}^{z}\rangle $.

In the mean-field approximation, the subsequent procedure amounts to
decoupling the chain of the equations of motion at the second step using the
triple-site term splitting of the characteristic form
\begin{equation}
\begin{array}{c}
S_{\mathbf{n+g}_{1}+\mathbf{g}_{2}}^{j}S_{\mathbf{n+g}_{1}}^{l}S_{\mathbf{n}}^
{\gamma }\approx \alpha _{\mathbf{g}}(\delta _{jl}\left\langle
S_{\mathbf{n+g}_{1}+\mathbf{g}_{2}}^{j}S_{\mathbf{n+g}_{1}}^{l}\right\rangle
S_{\mathbf{n}}^{\gamma }+ \\
+\delta _{l\gamma }\left\langle
S_{\mathbf{n+g}_{1}}^{l}S_{\mathbf{n}}^{\gamma }\right\rangle S_{\mathbf{n+g}_{1}+\mathbf{g}_{2}}^{j})+\alpha
_{\mathbf{g}_{1}+\mathbf{g}_{2}}\delta _{j\gamma }\left\langle
S_{\mathbf{n+g}_{1}+\mathbf{g}_{2}}^{j}S_{\mathbf{n}}^{\gamma }\right\rangle
S_{\mathbf{n+g}_{1}}^{l}\end{array}
\label{a_deco}
\end{equation}
where $\alpha _{\mathbf{g}}$ and $\alpha
_{\mathbf{g}_{1}+\mathbf{g}_{2}}$ are vertex corrections. In the
general case vertex corrections can depend on cite indices,
hereafter we use the simplest one-vertex approximation
\cite{Kondo72_PTP,Shimah91_JPSJ,Hartel10_PRB,Mikhey13_JL}, i.e.
all the vertices are taken to be equal. We emphasize that in the
case of $S=1/2$ the splitting procedure for each term is
unambiguous because in (\ref{a_Eq2_GF}) $\beta \neq z$, and the
average involves two spins with $\beta \neq z$.

After the Fourier transformation
\begin{equation}
S_{\mathbf{q}}^{z}=\frac{1}{\sqrt{N}}{\,}\sum\limits_{\mathbf{r}}
e^{-i\mathbf{qr}}S_{\mathbf{r}}^{z}
\end{equation}
solution of equations (\ref{a_Eq1_GF}) and (\ref{a_Eq2_GF}) leads
to the Green's function $G(\mathbf{q},\omega )=\langle
S_{\mathbf{q}}^{z}|S_{-\mathbf{q}}^{z}\rangle _{\omega }=-\chi
(\mathbf{q},\omega )$
\begin{equation}
G(\mathbf{q},\omega )=\frac{F_{\mathbf{q}}}{(\omega ^{2}-\omega
_{\mathbf{q}}^{2})}  \label{a_GFmf1}
\end{equation}

In the case of $J_{1}--J_{2}$ model the expressions for $F_{\mathbf{q}}$
and $\omega _{\mathbf{q}}^{2}$ are the following:
\begin{equation}
F_{\mathbf{q}}=-8\left[ J_{1}c_{g}(1-\gamma _{g})+J_{2}c_{d}(1-\gamma
_{d})\right] ;  \label{a_Fq2}
\end{equation} \begin{equation}
\omega _{\mathbf{q}}^{2}=2[(\gamma _{1}K_{1}+\gamma
_{2}K_{2})-(\gamma _{3}K_{3}+\gamma _{4}K_{4})-(\gamma
_{5}K_{5}+\gamma _{6}K_{6})] \label{a_w2q_1}
\end{equation}

The variables $K_{1}...K_{6}$ involved in the expression for the spectrum
are given by
\begin{equation}
K_{1}=J_{1}J_{2}K_{gd}+12J_{1}^{2}\widetilde{c}_{g}+1+K_{gg};\quad
K_{2}=J_{1}J_{2}K_{gd}+12J_{2}^{2}\widetilde{c}_{d}+1+K_{dd}
\end{equation}\begin{equation}
K_{3}=16J_{1}^{2}\widetilde{c}_{g};\quad
K_{4}=16J_{2}^{2}\widetilde{c}_{d};\quad K_{5}=16J_{1}J_{2}\widetilde{c}_{g};\quad
K_{6}=16J_{1}J_{2}\widetilde{c}_{d}  \label{a_K1-6}
\end{equation}\begin{equation}
K_{gg}=\sum_{\substack{\mathbf{r=g}_{1}\mathbf{+g}_{2}; \cr \mathbf{g}_{1}\neq
-\mathbf{g}_{2}}}\widetilde{c}_{r};\
K_{dd}=\sum_{\substack{\mathbf{r=d}_{1}\mathbf{+d}_{2}; \cr \mathbf{d}_{1}\neq
-\mathbf{d}_{2}}}\widetilde{c}_{r};\
K_{gd}=\sum_{\mathbf{r=g+d}}\widetilde{c}_{r}
\end{equation}\begin{eqnarray}
\gamma _{1} &=&1-\gamma _{g}; \, \gamma _{2}=1-\gamma _{d}; \, \gamma
_{3}=1-\gamma _{g}^{2}; \, \gamma _{4}=1-\gamma _{d}^{2}; \, \gamma
_{5}=(1-\gamma _{g})\gamma _{d}; \\
\gamma _{6} &=&(1-\gamma _{d})\gamma _{g};\quad \gamma _{g}=\frac{1}{2}(\cos
q_{x}+\cos q_{y});\quad \gamma _{d}=\cos q_{x}\cos q_{y}
\end{eqnarray}
where $\widetilde{c}_{r}$ stands for the correlators, renormalized
by the vertex correction $\ \widetilde{c}_{r}=\alpha c_{r}$ (in
(\ref{a_w2q_1}) we rearranged the contributions to the spectrum in
a different manner, than in \cite{Hartel10_PRB,Hartel13_PRB}).

The Green's function $G(\mathbf{q},\omega )$ involves the
correlators $c_{\mathbf{r}}$ for the first five coordination spheres $\mathbf{r}=\mathbf{g},\mathbf{d},2\mathbf{g},\mathbf{g+d},2\mathbf{d}$,
which must be evaluated self-consistently in terms of $G(\mathbf{q},\omega
)$. In addition, $G(\mathbf{q},\omega )$ must satisfy the spin
constraint $c_{\mathbf{r=0}}=\left\langle
S_{\mathbf{i}}^{z}S_{\mathbf{i}}^{z}\right\rangle =1/4$\ (sum
rule). These conditions are
\begin{equation}
c_{\mathbf{r}}=\frac{1}{N}\sum_{\mathbf{q}}c_{\mathbf{q}}e^{i\mathbf{qr}}; \, c_{\mathbf{q}}=\left\langle
S_{\mathbf{q}}^{z}S_{\mathbf{-q}}^{z}\right\rangle =-\frac{1}{\pi}
\int_{0}^{\infty }d\omega \,\left(
2m(\omega )+1\right) \mathrm{Im}G^{zz}\left( \omega ,\mathbf{q}\right) ;\
\label{a_Cq_1}
\end{equation} \begin{equation}
c_{\mathbf{r=0}}=1/4\ =-\frac{1}{\pi
}\frac{1}{N}\sum_{\mathbf{q}}\int_{0}^{\infty }d\omega \,\left( 2m(\omega )+1\right)
\mathrm{Im}G^{zz}\left( \omega ,\mathbf{q}\right) ;  \label{a_constr_1}
\end{equation}

The system of self-consistent equations is then solved numerically.
Hereafter all the energy-related parameters are set in the units of
$J=\sqrt{J_{1}^{2}+J_{2}^{2}}$.

In the general case, in the framework of (\ref{a_Eq1_GF}), (\ref{a_Eq2_GF}),
both short-range order (SRO) and long-range order (LRO) states can be
realized. Because the dimension is equal to two, only SRO are possible at
$T\neq 0$, and both possibilities can take place as $T\rightarrow 0$ (LRO is
characterized by nonzero spin--spin correlators at infinity $\langle
S_{\mathbf{r}}^{\alpha }S_{\mathbf{0}}^{\alpha }\rangle _{r\rightarrow \infty}$).

\section{The ground state properties}
\label{sec:GroSt}

For $T\neq 0$ the system is always in the SRO state (the
spin-liquid state). For small $T$, the function $c_{\mathbf{q}}$
is always peaked near the point $\mathbf{q}_{0} $ where the spin
gap is minimum (except for the zero point $\mathbf{q}=0 $, where
spin gap is always zero, but $c_{\mathbf{q}}$ is not peaked
because $F_{\mathbf{q=0}}=0$).

Two cases are possible as $T\rightarrow 0$. In the first case, the
system remains in the SRO state, the gap is not closed, and the
contribution of $m(\omega )$ to $c_{\mathbf{q}}$ in (\ref{a_Cq_1})
vanishes as $T\rightarrow 0$.

In the second case, as the temperature decreases, the system
passes into the LRO state, and the gap goes to zero at a point
$\mathbf{q}_{0}$, which for $J_{1}-J_{2}$ model can be either
$\mathbf{q}_{0}=\mathbf{Q}=(\pi ,\pi )$, (corresponding to
antiferomagnetic LRO) or $\mathbf{q}_{0}=\mathbf{X}=(\pi ,0);\
(0,\pi )$ (stripe LRO). The zero gap leads to the existence of
condensate $c_{cond}$, which determines spin-spin correlation
function at infinity. That is nonzero condensate means the LRO
existence at zero temperature.

The case of FM LRO is somewhat different. The condensate at the
point $\mathbf{q}_{0}=\mathbf{\Gamma }=(0,0)$ does appear for $T\rightarrow
0 $, but the spin gap at $\mathbf{q}_{0}$\ is closed for any temperature. In
this case the transition to LRO is governed by the spectrum transformation
near $\mathbf{q}_{0}$ --- from linear to quadratic (see detailes in
\cite{Mikhey13_JL}).

Note that the third exchange interaction $J_{3}$ being added to the model
changes the LRO\ picture qualitatively --- the helical LRO becomes possible.
In the $J_{1}-J_{2}-J_{3}$ model the condensate peak point in the structure
factor can be located not only at $\mathbf{\Gamma }$, $\mathbf{Q}$, or
$\mathbf{X}$, but also at arbitrary incommensurate point on the side or
diagonal of the Brillouin zone
\cite{Baraba11_TMP,Sindzi09_JPCS,Mikhey11_JL,Mikhey12_SSC,Sindzi10_JPCS}.

\begin{figure}[tbp]
\begin{center}
\includegraphics[width=.8\textwidth]{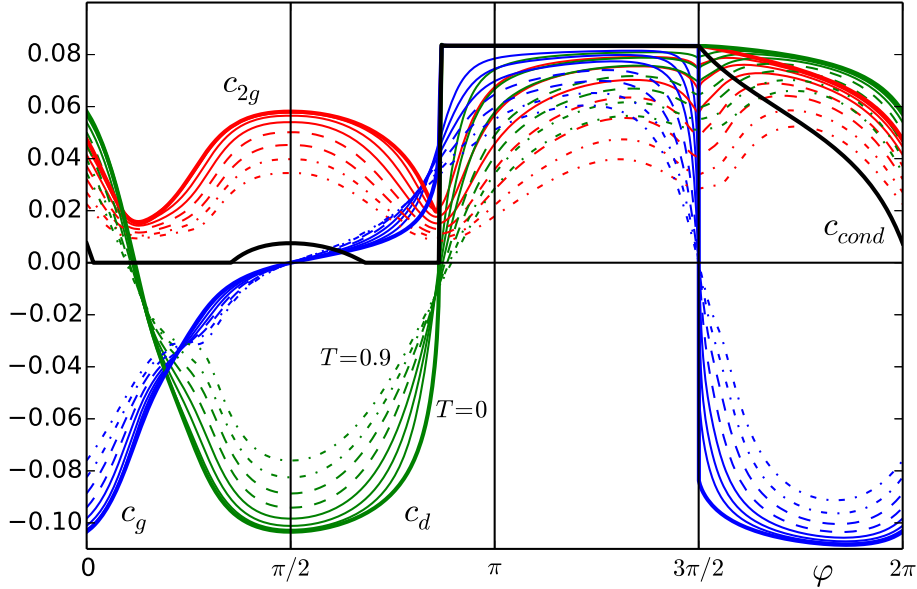}
\caption{(Color online) Condensate $c_{cond}$ (black bold line)
and correlation functions on the first three coordination spheres
as functions of the angle parameter $\varphi$ ($J_1=\cos \varphi$,
$J_2=\sin \varphi$) --- $c_g$ (blue), $c_d$ (green) and $c_{2g}$
(red); bold lines correspond to the zero temperature, solid lines
--- to $T=0.3,0.4,0.5$, dashed lines --- to $T=0.6,0.7$ and dotted
lines --- to $T=0.8,0.9$. } \label{fig:fig2}
\end{center}
\end{figure}

The resulting picture for the whole region of $J_{1}$ and$\ J_{2}$ exchanges ("$J_{1}$--$J_{2}$
circle") \cite{Mikhey13_JL} is shown in Figures~\ref{fig:fig2},\ref{fig:fig4}. As seen in the
figures, four types of the ground state are possible: AFM, stripe, FM and the disordered
spin-liquid state (two different areas --- SL$^{1}$ and SL$^{2}$). Note, that in the frames of SSSA
the transition SL$^{2}\rightarrow $FM appears to be continuous (\cite{Mikhey13_JL}).

Let us remind, that in SSSA mean spin projection is always zero $\langle
S_{\mathbf{n}}^{z}\rangle =0$ and any possible ground state preserves the whole
--- translational and spin SU(2) --- symmetry of the Hamiltonian. The
LRO, if it does exist (this possibility is open --- LRO is possible
only at $T=0$) is determined by the spin-spin correlation function at
infinity.

At $\varphi =0$ ($J_{1}=1$, $J_{2}=0$), as it the classical limit,
the ground state is AFM. Quantum fluctuations destroy LRO with the
increasing $\varphi $ (that is increasing $J_{2}$) and the system
transforms to the disordered state SL$^{1}$. Note, that different
alternative states are considered to be competitive in the
locality of LRO disappearance at $T=0$. These are in particular
columnar and box phases which preserve the SU(2) symmetry, but
brake the translational one (see \cite{Sandvi10_AIP} for recent
review). The disordered state in SSSA is always spin-liquid-like
in the above noted sense and it can not be distinguished from the
mentioned alternatives with nearby energies.

\begin{figure}[h]
\begin{center}
\begin{minipage}{17pc}
\includegraphics[width=17pc, trim=2pc 3pc 2pc 0]{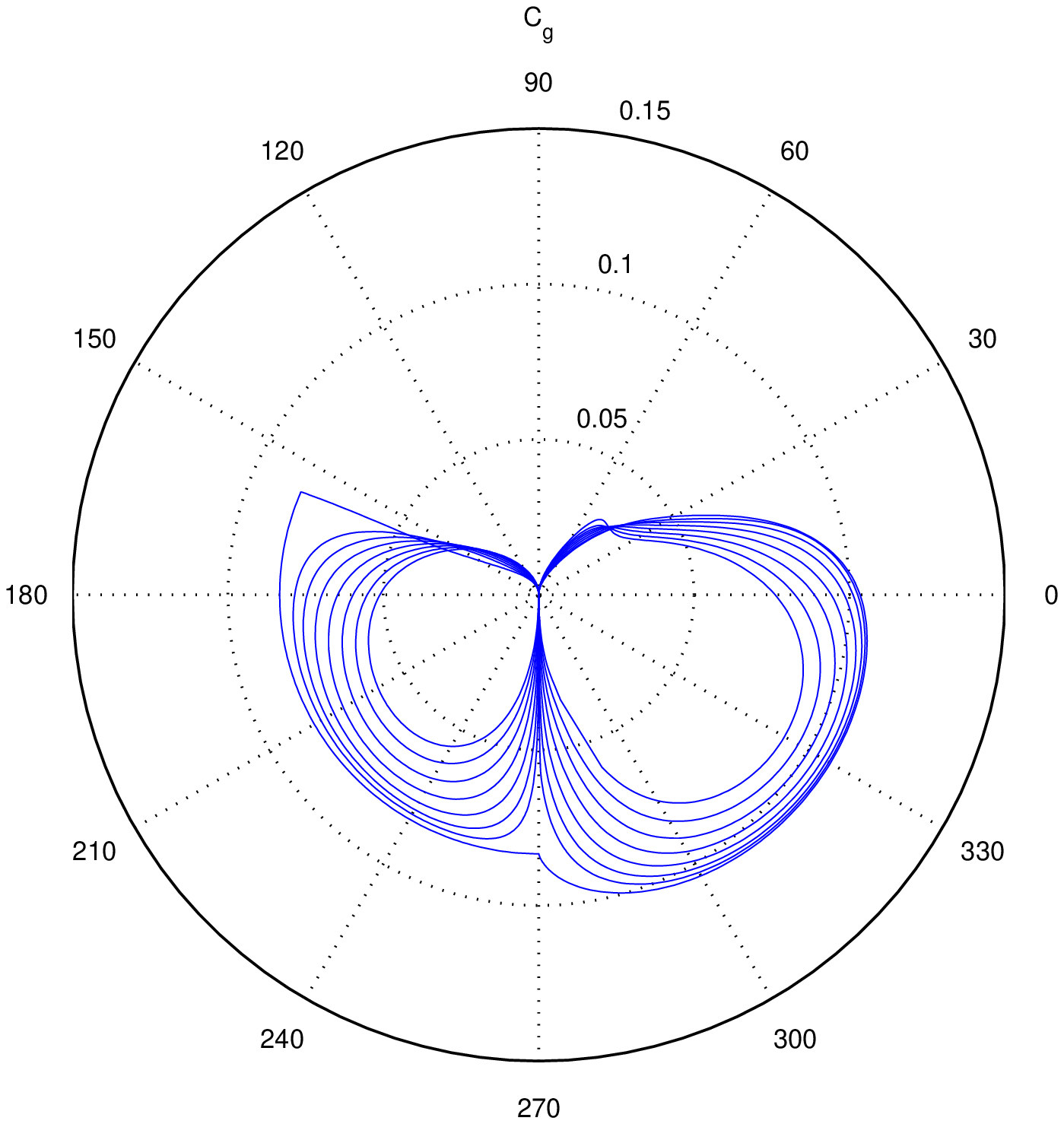} \\ \hspace*{3cm} a
\end{minipage}\hspace{.5pc}
\begin{minipage}{17pc}
\includegraphics[width=17pc, trim=2pc 3pc 2pc 0]{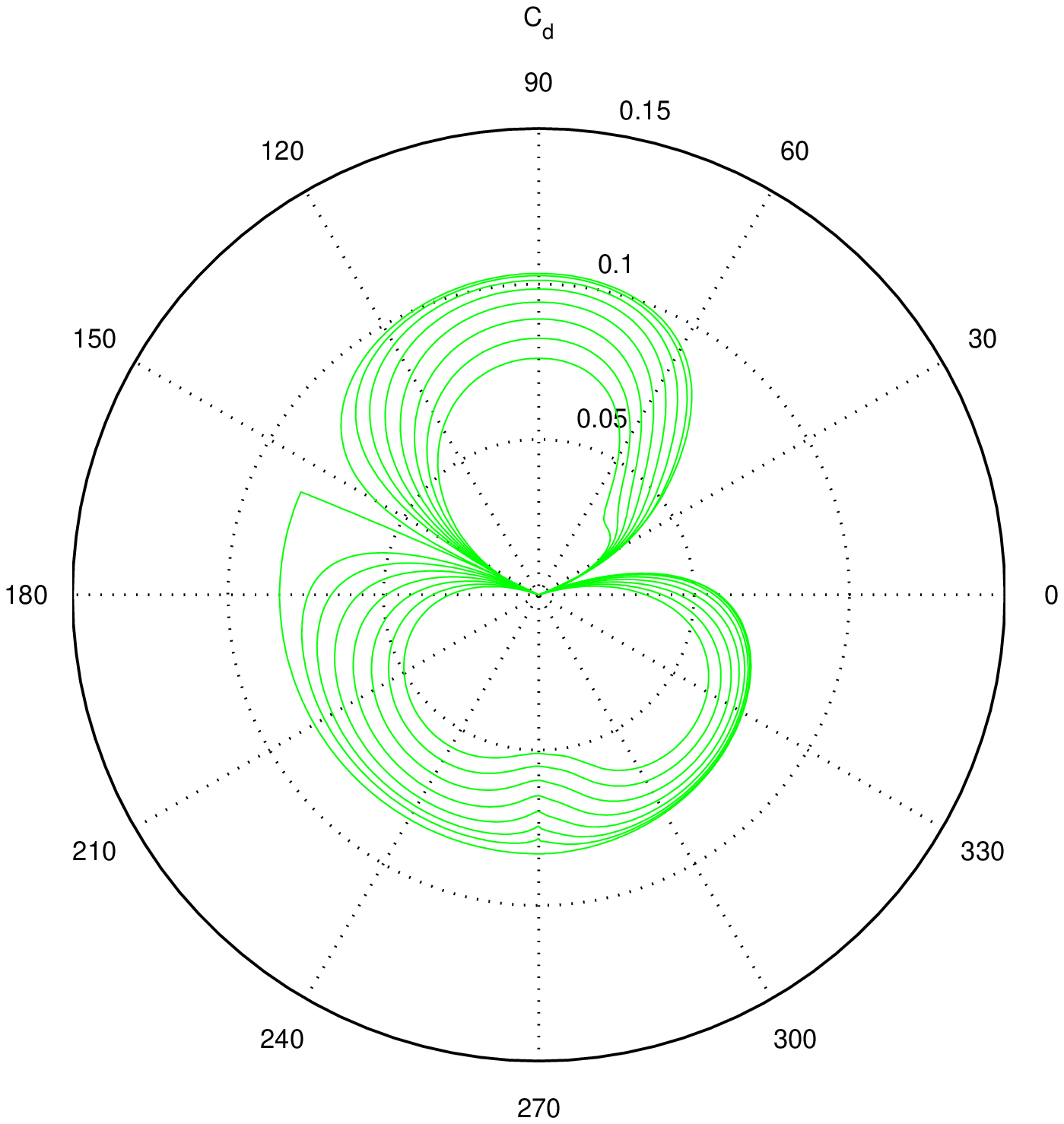} \\ \hspace*{3cm} b
\end{minipage}\hspace{.5pc}
\begin{minipage}{17pc}
\includegraphics[width=17pc, trim=2pc 3pc 2pc 0]{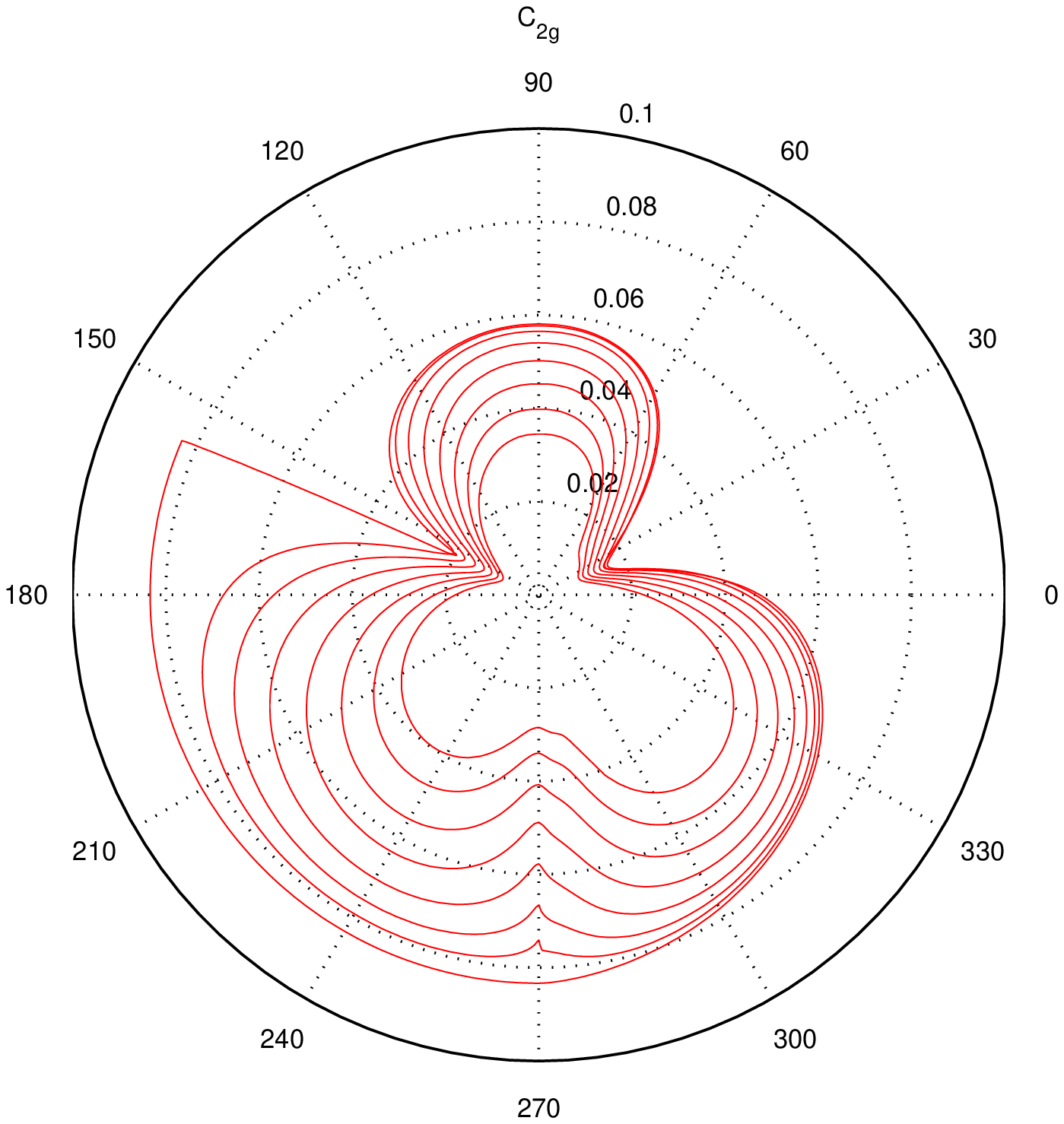} \\ \hspace*{3cm} c
\end{minipage}
\caption{(Color online) Polar diagrams for the absolute values of
the first (a), second (b) and third (c) correlation functions
($|c_g|$, $|c_d|$, and $|c_{2g}|$ correspondingly); outer lines
correspond to $T=0$, inner lines --- to $T=0.3 \div 0.9$.}
\label{fig:fig3}
\end{center}
\end{figure}

In the vicinity of $\varphi =\pi /2$ there exists another ordered
state, with stripe LRO type. Then with the increasing $\varphi $
the spin liquid appears again (SL$^{2}$, its SRO differing from
that of SL$^{1}$). After SL$^{2}$ the system continuously
transforms to FM-LRO state. And the last transition --- at
$\varphi =3\pi /4$ restores the AFM LRO. In the frames of SSSA all
the mentioned transitions at $T=0$ are continuous, except the last
one.

\section{Thermodynamic properties}
\label{sec:Thermod}

The results of our calculations for the spin-spin correlation finctions on the first, second and
third coordination spheres ($c_{\mathbf{g}}$, $c_{\mathbf{d}}$ and $c_{2\mathbf{g}}$
correspondingly) at different temperatures ($0\leq T\leq 0.8$) are presented in
Figure~\ref{fig:fig2}.

As it is seen from the figure, at nonzero temperatures the SRO of the
disordered state holds the memory of the parent zero-temperature ordered
phase. The region overlying the AFM phase is characterized by AFM-like
correlators $c_{g}<0$, $|c_{g}|>c_{d}>c_{2g}>0$, the region above stripe
phase -- by stripe-like ones $c_{d}<0,$ $c_{2g}>0$, $|c_{d}|>c_{2g}>|c_{g}|$,
and the area above FM phase -- by FM-like $c_{g}>c_{d}>c_{2g}>0$. In the
intermediate regions SRO transforms from one limit to another.

\begin{figure}[tbp]
\begin{center}
\includegraphics[width=.8\textwidth]{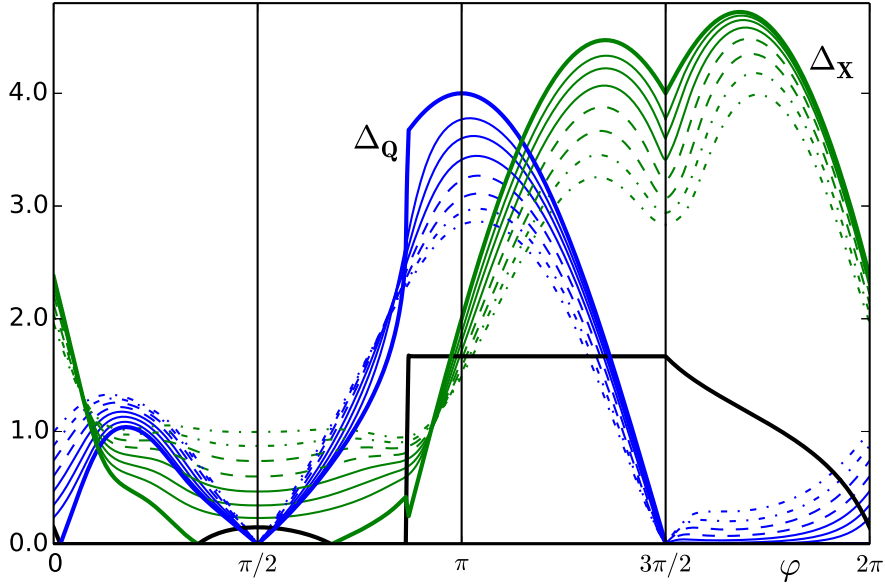}
\caption{(Color online) Spin gaps $\Delta_\mathbf{Q}$ (blue) and
$\Delta_\mathbf{X}$ (green) at points $\mathbf{Q}=(\pi,\pi)$ and
$\mathbf{X}=(\pi,0),(0,\pi)$ as functions of the angle parameter
$\varphi$ ($J_1=\cos \varphi$, $J_2=\sin \varphi$); bold lines
correspond to the zero temperature, solid lines --- to
$T=0.3,0.4,0.5$, dashed lines --- to $T=0.6,0.7$ and dotted lines
--- to $T=0.8,0.9$. Black bold line corresponds to the value of
condensate function $c_{cond}$ (in altered scale). }
\label{fig:fig4}
\end{center}
\end{figure}

It is natural to expect simultaneous decrease of all the
correlators absolute values with the temperature growth. In the
main so it is. There are, however, two important exceptions with
inverse temperature dependence. It is firstly the nearest
correlator $c_{g}$ in the region of small $J_{1}$ -- above the
stripe phase, in part of SL$^{1}$ and in the whole SL$^{2}$
region; secondly, NN correlator $c_{d}$ in narrow vicinities of
the nodes -- in the middle of SL$^{1}$ and near the transition
SL$^{2}\rightarrow $FM (see Sec.~\ref{sec:Diss} for the explanations).

The general view of the correlators temperature behaviour can be also seen in
Figure~\ref{fig:fig3}a, Figure~\ref{fig:fig3}b and Figure~\ref{fig:fig3}c, where polar diagrams for
the correlators absolute values are shown. These figures allow to compare the SRO structure with
the initial classical phase diagram. In particular, it is clearly seen, that in the region
corresponding to classical stripe phase the nearest correlator $|c_{g}|$ is an order of magnitude
smaller than the next-nearest one $|c_{d}|$. The regions of $c_{g}$ and $c_{d}$ inverse temperature
dependence can be also seen.

\begin{figure}[h]
\begin{center}
\begin{minipage}{17pc}
\includegraphics[width=17pc, trim=2pc 0 2pc 0]{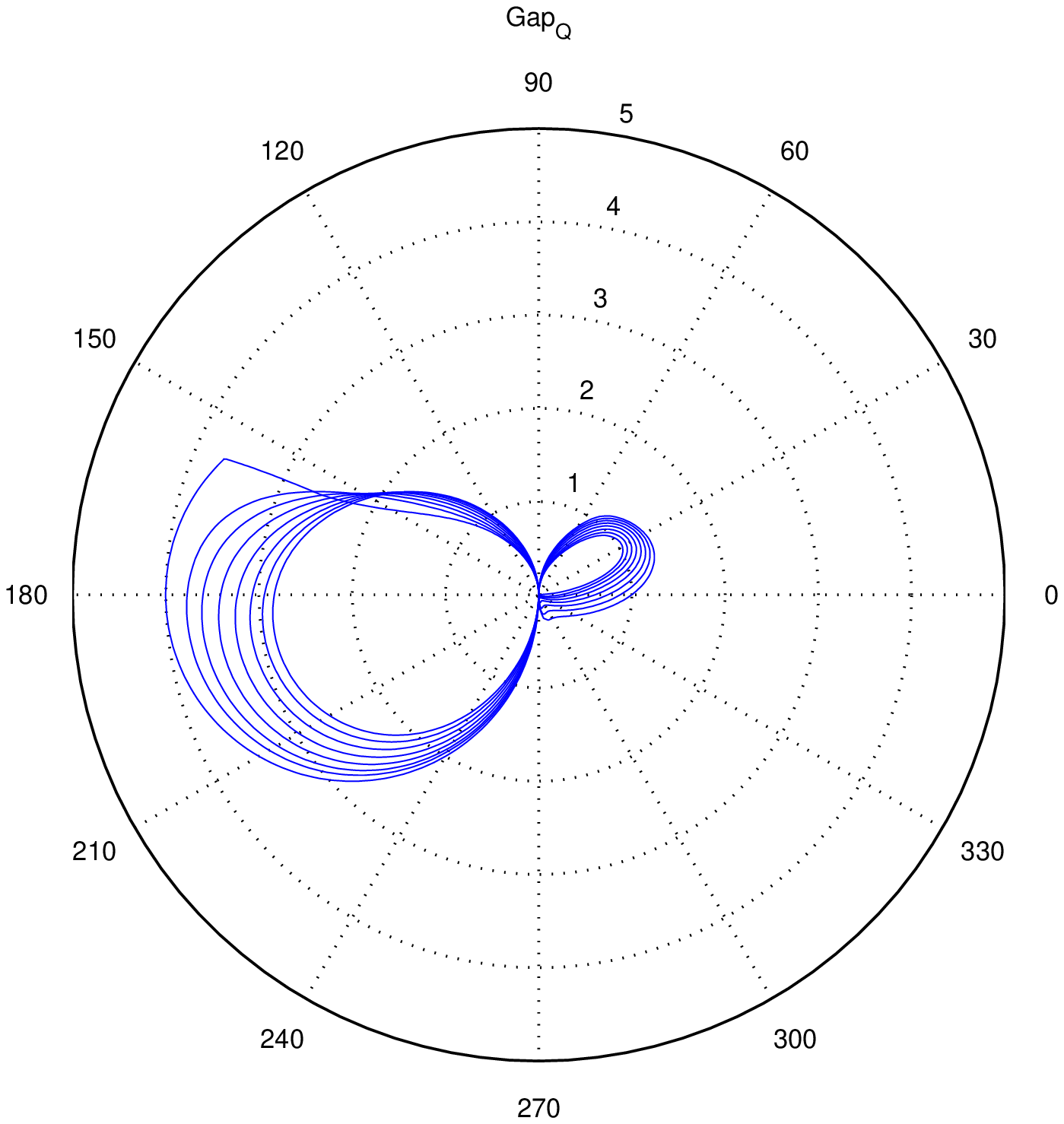} \\ \hspace*{3cm} a
\end{minipage}\hspace{1pc}
\begin{minipage}{17pc}
\includegraphics[width=17pc, trim=2pc 0 2pc 0]{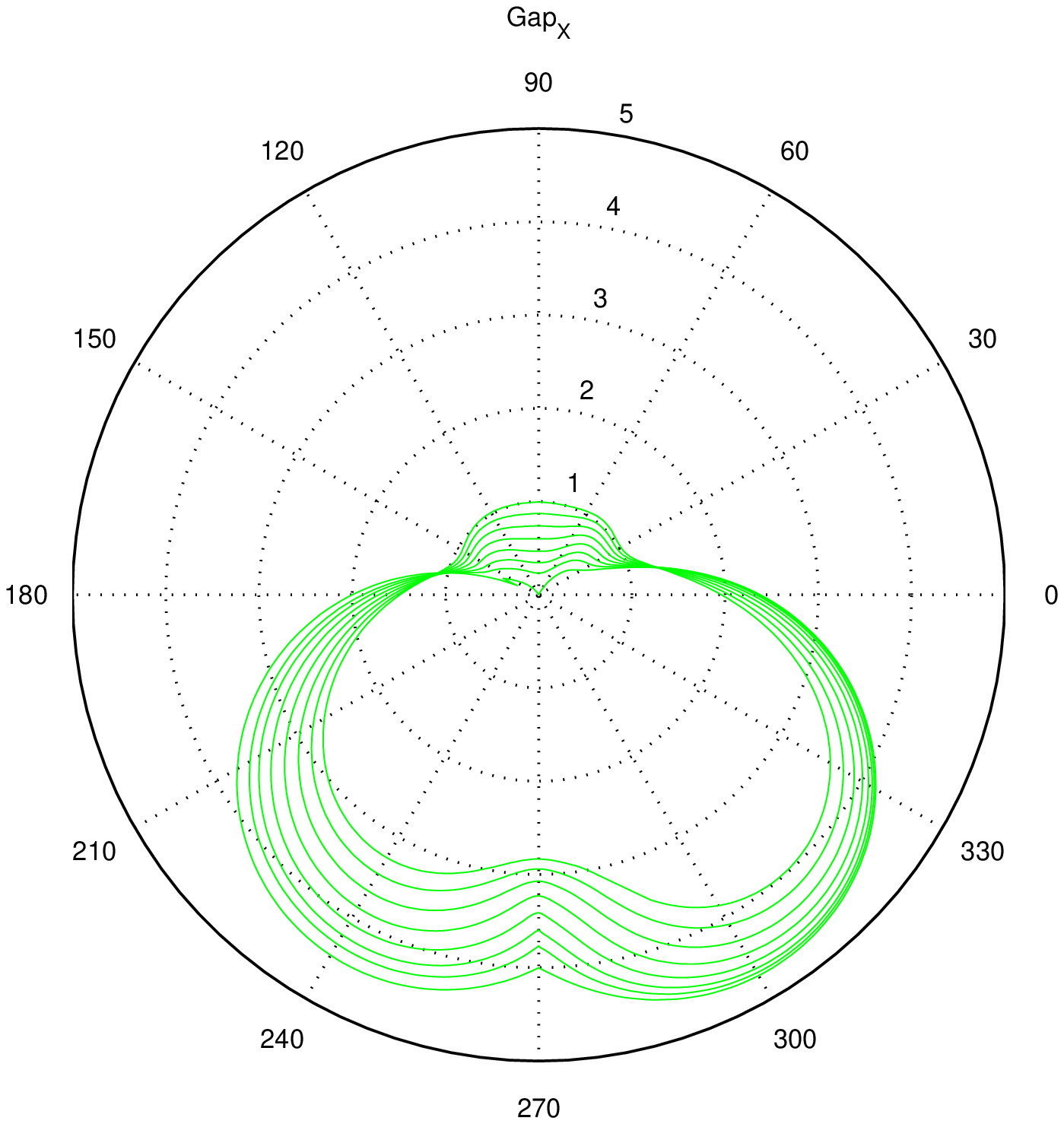} \\ \hspace*{3cm} b
\end{minipage}
\caption{(Color online) Polar diagrams for the spin gaps
$\Delta_\mathbf{Q}$ (a) and $\Delta_\mathbf{X}$ (b) (compare with
Fig.\ref{fig:fig4}).} \label{fig:fig5}
\end{center}
\end{figure}

The gap in the spin excitations spectrum at zero point
$\mathbf{\Gamma =}(0,0)$ of the Brillouin zone is zero $\Delta
_{\mathbf{\Gamma }}=0$ in any phase at any temperature. The
temperature dependence of spin gaps $\Delta _{\mathbf{Q}}$ and
$\Delta _{\mathbf{X}}$ at two another symmetrical points --- AFM
point $\mathbf{Q}=\left(\pi ,\pi \right)$ and (equivalent) stripe
points $\mathbf{X}=(0,\pi ),(\pi ,0)$ --- is shown in
Fig.~\ref{fig:fig4} (the corresponding polar diagrams --
Fig.~\ref{fig:fig5}a è Fig.~\ref{fig:fig5}b).

Figure~\ref{fig:fig6} shows the results for heat capacity $C_{V}(\varphi )$ at different
temperatures. Two next figures --- Figures~\ref{fig:fig7}, \ref{fig:fig8} --- show the detailed
temperature dependencies for several values of $\varphi$ (in the temperature intervals, where the
adopted computational scheme leads to satisfactory convergence). The two mentioned figures
demonstrate in particular, that the temperature curve for heat capacity has a peak for any chosen
value of $\varphi$.

\section{Discussion and summary}
\label{sec:Diss}

Hereinafter we discuss the most interesting properties of  the phase diagram obtained.
Let us remind, that the LRO is absent at $T\neq 0$ and all  the "circle"\
$0\leq \varphi \leq 2\pi $ is covered by spin-liquid. Nevertheless we will classify
the $T\neq 0$ phase diagram regions by the LRO type at $T=0$.

\subsection{AFM order}

\subsubsection{} A dramatic difference of the $T=0$ LRO strength
(the value of spin-spin correlator at infinity) at points
$\varphi =0$ $(J_{1}=1$, $J_{2}=0)$ and $\varphi =3\pi
/2+0$\ $(J_{1}=+0$, $J_{2}=-1)$ is worth noting.

That is the infinitesimal AFM NN exchange with the FM NNN (diagonal) exchange
equal to unity leads to much more strong LRO than in the case
of conventional AFM with NN AFM exchange equal to unity. The SRO
difference between these two points (both at zero and nonzero temperatures) is not
so large.

\subsubsection{} In the interval between $\varphi =3\pi /2+0$ and
$\varphi \sim 330^{\circ }$ (Figure~\ref{fig:fig4} and Figure~\ref{fig:fig5}a) the AFM spin gap is
small ($\Delta _{\mathbf{Q}}\ll T$) up to high enough temperatures ($T\lesssim 0.5$). At the
conventional AFM point $\varphi =0$ the gap becomes exponentially small at much lower temperatures
($T\lesssim 0.05 \ll 0.5 $). When $\Delta _{\mathbf{Q}}\ll T$ the gap $\Delta _{\mathbf{Q}}$
determines the AFM correlation length $\xi _{AFM}\sim \Delta _{\mathbf{Q}}^{-1}$. So the
correlation length increases greatly from the point $\varphi =0$ to the interval $3\pi /2\leq
\varphi \lesssim 330^{\circ }$ (that is consistent with the mentioned LRO evolution).

Note, that, as it can be shown, at $\varphi =0$ the
exponentially small gap at high temperatures $T\sim 0.5$ is
realized for the spin $S\geq 1$.

\subsection{Spin liquid}

\subsubsection{} Let us note the dissimilarity of the spin liquid
evolution (with growing $\varphi $) in two areas, corresponding to zero-temperature regions
SL$^{1}$ and SL$^{2}$. The SL$^{1}$ liquid at $T=0$ exists in the angle range $0.051 \le \varphi
\le 1.111$ ($2.92^{\circ} \le \varphi \le 57.30^{\circ}$), it is between AFM and stripe phases. The
SRO in SL$^{1}$ smoothly transforms from the left bank to the right one, from the AFM-type
($c_{g}<0$, $|c_{g}|>c_{d}>c_{2g}>0$) to the stripe-type ($c_{d}<0,$ $c_{2g}>0$,
$|c_{d}|>c_{2g}>|c_{g}|$). That is not the case for SL$^{2}$ liquid ($2.141 \le \varphi \le 2.712$
($122.67^{\circ} \le \varphi \le 155.39^{\circ}$) at $T=0$, the region between stripe and FM
phases). Almost in the whole SL$^{2}$ area the SRO is stripe-like, in particular, $c_{d}$ remains
negative. The absolute value of $c_{d}$ almost everywhere, except tiny region near $\varphi =
2.712$, is larger than the nearest neighbour correlator $c_{g}$. And only close to FM border the
drastic restructuring of the correlators takes place. At $T\neq 0$, as it was repeatedly mentioned,
there is no LRO, but all the above statements concerning SRO do hold.

\subsubsection{} As it was noted earlier, there are regions of  the phase
diagram with anomalous temperature behaviour of the correlators
at fixed $\varphi $ (absolute value
growing with temperature or nonmonotonic behaviour).

This anomaly for $c_{g}$ correlator realizes in the region from
the middle of SL$^{1}$ phase through the stripe-phase and all the
SL$^{2}$ up to the transition to FM. The anomalous regions for
$c_{d}$ correlator --- narrow areas near the nodes --- are in the
middle of SL$^{1}$ and near SL$^{2}\rightarrow $FM transition. The
reason for $c_{d}$ anomaly is obviously the following. For
different temperatures $c_{d}(\varphi )$ changes sign at different
points $\varphi $. The normal temperature behaviour of the
correlator would pass to the normal one after crossing $\varphi$
axis only if the node of the correlators cone should be exactly on
the $\varphi$ axis (that is $c_{d}(\varphi _{0},T)=0$ for any
$T$). But generally there is no physical reason for this
statement.

The reasons for other mentioned temperature anomaly --- the
$c_{g}(\varphi ,T)$ behaviour --- are not so obvious. Presumably
it is connected to the rapid SRO rearrangement.

\subsection{Stripe order}

The most interesting point of the area is $\varphi=\pi /2$ ($J_{1}=0$, $J_{2}=1$). At this point
the lattice is decoupled into two non-interacting sublattices. One can see from
Figure~\ref{fig:fig2}, that, as it should be, at any temperature $c_{d}\left( \pi /2\right)
=c_{g}\left( 0\right) $, $c_{2g}\left( \pi /2\right) =c_{d}\left( 0\right) $. The decoupling means
that $c_{g}(\pi /2,T)$ is strictly zero. That is why the cone of $c_{g}$ correlators with anomalous
$T$-behaviour retains the anomalous behaviour after crossing the node on $\varphi$ axis.

At the same point the AFM gap $\Delta _{\mathbf{Q}}(\varphi =\pi /2)=0$ for any temperature (see
Figure~\ref{fig:fig4}), though AFM LRO at any temperatures, including $T=0$, is absent.

Formally it follows from the analytical expression for the AFM gap
$\Delta _{\mathbf{Q}}(\varphi =\pi /2)\sim J_{1}$. The naive
explanation is that the system is degenerate with respect to
mutual rotation of the sublattices, that is the transfer of spin
excitation to the neighbouring cite costs no energy.

All the same is true for the point $\varphi =3\pi /2$ ($J_{1}=0$,
$J_{2}=-1$). Though is this case there is AFM LRO at $\varphi =3\pi /2+0$
and $T=0$ (and FM LRO at $\varphi =3\pi /2-0$, $T=0$).

\subsection{FM order}
\subsubsection{SL$^{2}\rightarrow $ FM transition}

As it was noted earlier, it was shown in \cite{Mikhey13_JL}, that SL$^{2}\rightarrow $ FM
transition at $T=0$ is continuous, though it takes place in the very narrow $\varphi$ interval. At
$T\neq 0$, as it is seen in Figures~\ref{fig:fig2}--\ref{fig:fig5}, in the vicinity of this
transition the $c_{2g}$ temperature dependence is nonmonotonic and the temperature dependencies of
other correlators and the gaps are inverse. The heat capacity (as function of $\varphi$) at any
temperature has sharp minimum near this transition (Figures~\ref{fig:fig6}--\ref{fig:fig8}).

\subsubsection{FM $\rightarrow $ AFM transition}

This transition takes place at $\varphi=3\pi /2$. At this point the lattice
is splitted into two noninteracting sublattices. At $\varphi \rightarrow 3\pi /2-0$
there is no frustration with respect to the FM order, at $\varphi \rightarrow 3\pi /2+0$
--- no frustration with respect to the AFM order. Therefore it is physically
obvious that in the quantum limit at $T=0$ a transition between these two phases is of the first
order and there is no spin-liquid area between FM an AFM. Our previous calculations
\cite{Mikhey13_JL} do confirm this evidence. At nonzero temperatures the transition is obviously
continuous. But as it is seen from Figures~\ref{fig:fig2}--\ref{fig:fig5}, the correlator
$c_{g}(\varphi)$ rapidly transforms and changes the sign. Other correlators and the spin gaps
demonstrate nonmonotonic dependence on $\varphi$, the heat capacity has sharp minimum.

Note once more, that near $\varphi =270+\Delta \varphi$ the AFM
gap $\Delta _{\mathbf{Q}}$ is exponentially small up to $T\sim
0.5$. $\Delta _{\mathbf{Q}}\sim 10^{-4}-10^{-5}$ for $\Delta
\varphi \sim 5^{o}$. That is why our calculation can not reproduce
the fine structure of the correlators and the heat capacity for
$\varphi =270+\Delta \varphi$ at low temperatures.

\subsection{Specific heat}

\begin{figure}[tbp]
\begin{center}
\includegraphics[width=.8\textwidth]{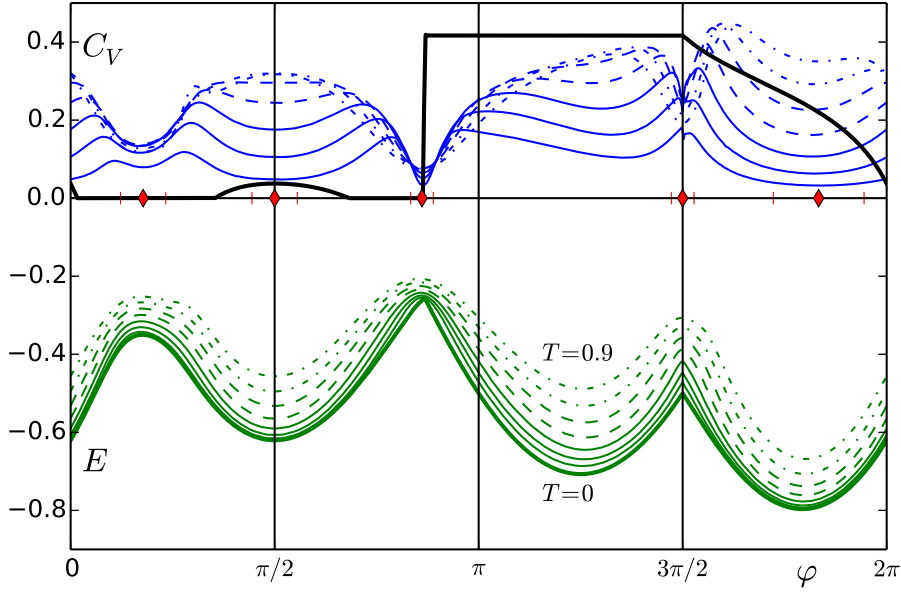}
\caption{(Color online) Specific heat $C_V$ (blue) and energy per site $E$ (green) as functions of
the angle parameter $\varphi$ ($J_1=\cos \varphi$, $J_2=\sin \varphi$). Bold green line corresponds
to the energy of the ground state ($T=0$). Solid lines correspond to $T=0.3,0.4,0.5$, dashed lines
--- to $T=0.6,0.7$ and dotted lines --- to $T=0.8,0.9$. Black bold line corresponds to the value of
condensate function $c_{cond}$ (in altered scale). The $C_V(T)$ slices for $\varphi$ values marked
by diamonds and tics are shown in Figures~\ref{fig:fig7}-\ref{fig:fig8}. } \label{fig:fig6}
\end{center}
\end{figure}

\begin{figure}[tbp]
\begin{center}
\includegraphics[width=.8\textwidth]{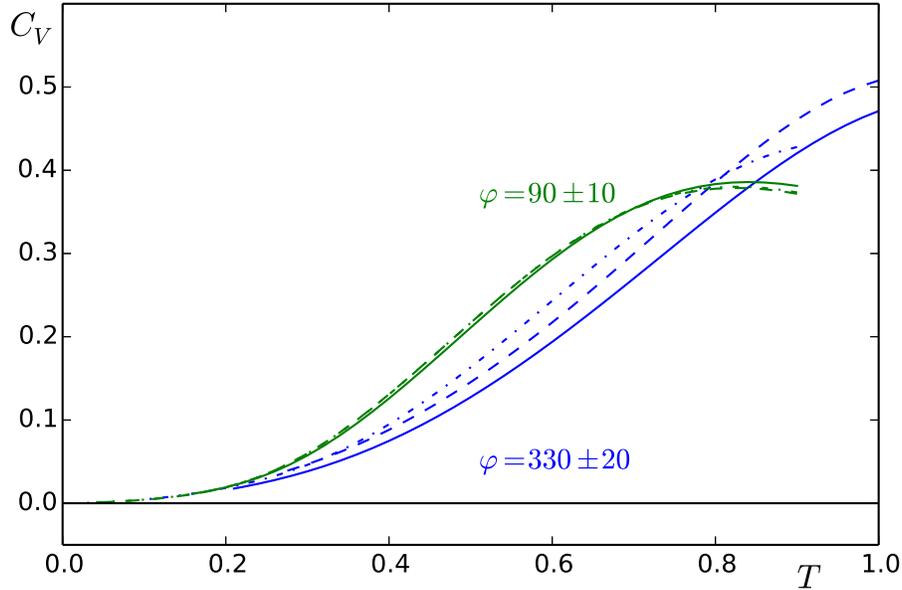}
\caption{(Color online) Families of the specific heat $C_V(T)$ curves near values of angle
parameter $\varphi$ ($J_1=\cos \varphi$, $J_2=\sin \varphi$) corresponding to $C_V(\varphi)$ minima
above FM and AFM phases (see Figure~\ref{fig:fig6}).} \label{fig:fig7}
\end{center}
\end{figure}

\begin{figure}[tbp]
\begin{center}
\includegraphics[width=.8\textwidth]{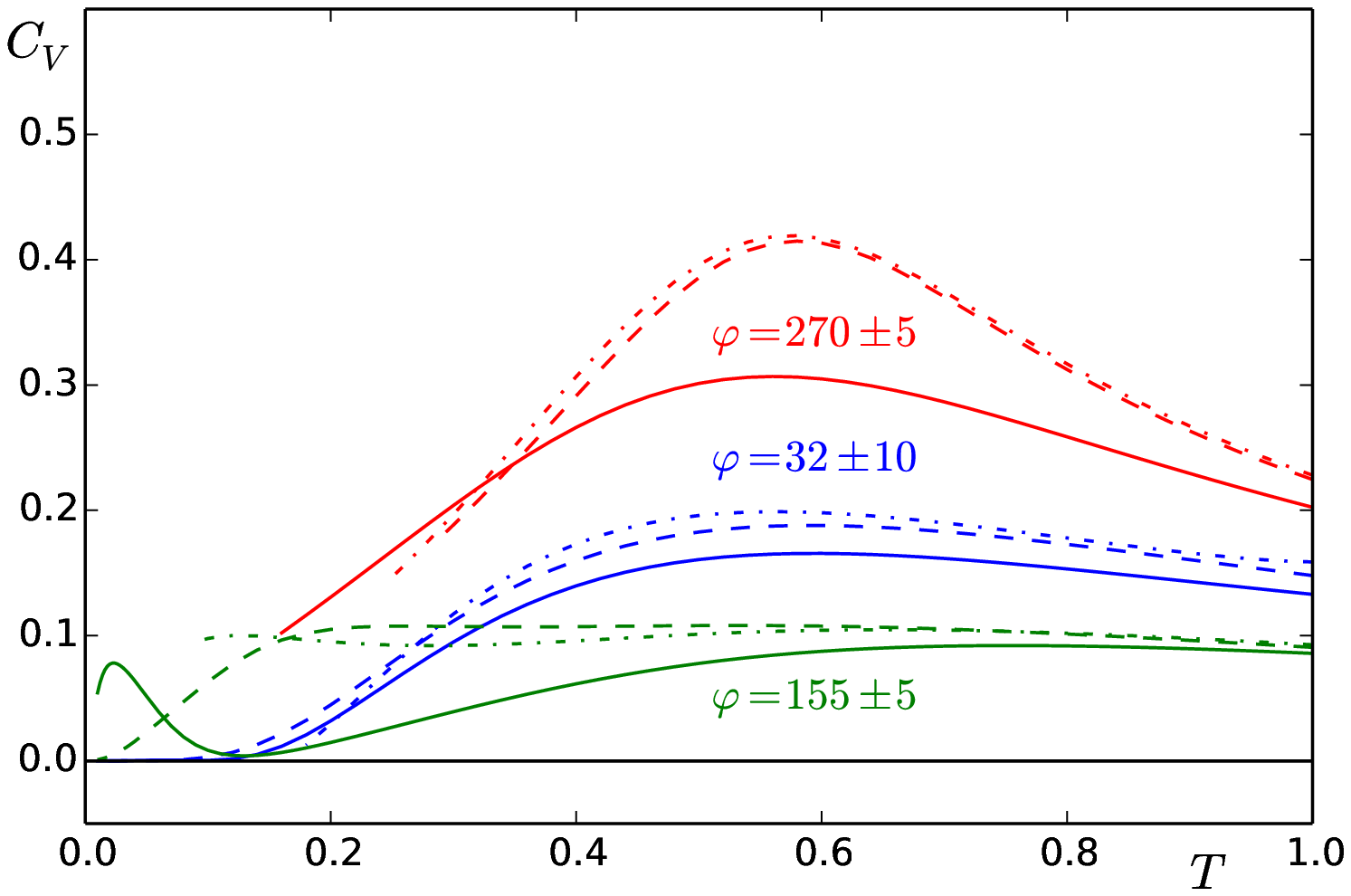}
\caption{(Color online) Families of the specific heat $C_V(T)$ curves near values of angle
parameter $\varphi$ ($J_1=\cos \varphi$, $J_2=\sin \varphi$) corresponding to $C_V(\varphi)$ minima
above the regions of SRO rearranging (see Figure~\ref{fig:fig6}). $C_V(T,\varphi=155^{\circ})$
demonstrates two maxima -- see text.} \label{fig:fig8}
\end{center}
\end{figure}

As it is seen from Figures~\ref{fig:fig6}--\ref{fig:fig8} the heat capacity for any $\varphi$ tends
to zero at $T\rightarrow 0$. The reason is the stabilization of all the correlators (and the
energy) at low temperatures. At any fixed $\varphi$ the heat capacity $T$-dependence, as it should,
has the maximum, varying in its height and position (see Figure~\ref{fig:fig7},
Figure~\ref{fig:fig8}).

It is also seen from Figure~\ref{fig:fig6} that there always exists a local minimum of the function
$C_{V}(\varphi )$ inside the areas above the ordered phases (AFM, stripe and FM), if $T$ is not too
high. The reason is that the correlators (and the energy) in the corresponding areas weakly depend
on $\varphi$. These minima correspond to local minima of the energy $E(\varphi)$.

Three other local minima of $C_{V}(\varphi )$ are observed, on the
contrary, in the regions of neighbouring short-range orders
rivalry, where the correlators are rapidly rearranging, and the
energy also weakly depends on $\varphi$. These minima correspond
to local maxima of the energy $E(\varphi)$.

Five of the mentioned points of $C_{V}(\varphi)$ local minima are marked by diamonds in
Figure~\ref{fig:fig6}. The heat capacity temperature dependencies $C_{V}(T)$ at these points and
symmetrically neighbouring points are shown in Figures~\ref{fig:fig7}, \ref{fig:fig8}.

Figure~\ref{fig:fig7} corresponds to $\varphi$ in the middle of ordered (at $T=0$) Stripe and AFM
phases. It is interesting that the heat capacity $C_{V}(\varphi,T)$ is symmetric on $\varphi$
relative to the minima of $C_{V}(\varphi)$ and $E(\varphi)$ in a wide $T$ range. The situation in
the region above the FM phase is slightly different, because, as it is seen from
Figure~\ref{fig:fig6}, $C_{V}(\varphi)$ moves with temperature.

Figure~\ref{fig:fig8} corresponds to $\varphi$ in the regions of the strong frustration where
different short-range orders concur. The heat capacity $C_{V}(\varphi,T)$ is symmetric on $\varphi$
relative to the minimum of $C_{V}(\varphi)$ and maximum $E(\varphi)$ in a wide $T$ range. The case
of $\varphi = 155^{\circ}\pm 5^{\circ}$ is obviously notable. As in other cases symmetrical
satellite $C_{V}(T)$ lines are similar, but the central line $C_{V}(\varphi=155^{\circ},T)$
demonstrates additional maximum at low $T$. This maximum was found and discussed in
\cite{Hartel10_PRB}. The low-temperature frustration-induced maximum $C_{V}(T)$ was also found in
1D case for $S=1/2$ and $S=1$ \cite{Hartel11_PRB}.

\subsection{Fine tuning of the method}

As it was noted above, all the calculations in the present work
has been carried out in the straight and simple approximation,
without any tuning parameters. The tuning in SSSA is commonly made
via different manipulations with vertex corrections
\cite{Baraba11_TMP,Juhasz08_PRB,Kondo72_PTP,Shimah91_JPSJ,Hartel10_PRB,
Hartel13_PRB,Hartel08_PRB,Vladim14_EPJB,Mikhey11_JL,Mikhey12_SSC,
Mikhey13_JL,Baraba94_JETP,Baraba94_JPSJ,Hartel11_PRB} or
accounting for complex structure of the Green's function, that is,
considering the polarization operator (in particular, accounting
for damping of spin excitations)
\cite{Baraba11_TMP,Mikhey11_JL,Mikhey12_SSC}.

\begin{figure}[tbp]
\begin{center}
\includegraphics[width=.7\textwidth]{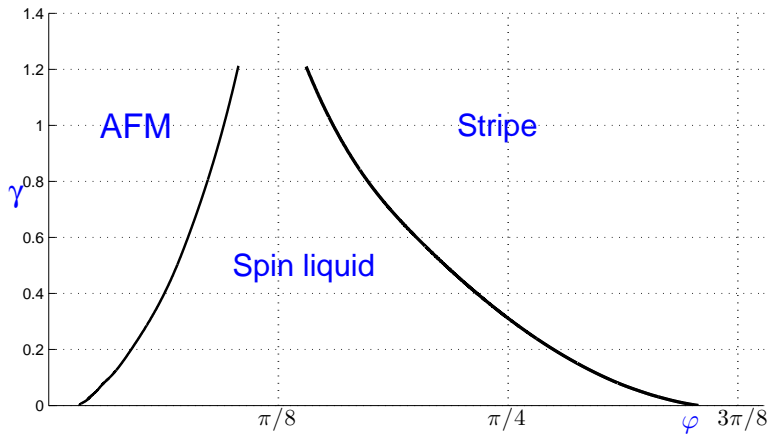}
\caption{(Color online)  Effect of the damping parameter $\gamma$
on the phase boundaries of the spin liquid SL$^{1}$.}
\label{fig:fig9}
\end{center}
\end{figure}

All such complications obviously affect the results. At very low $T$ the obtained quantitative
differences can amount significant values. Figure~\ref{fig:fig9} demonstrates, that even the
simplest self-consistent accounting for the damping in the frames of the Green's function
\begin{equation}
G_{\gamma }^{zz}\left( \omega ,\mathbf{q}\right)
=\frac{F_{\mathbf{q}}}{\omega ^{2}-\omega _{\mathbf{q}}^{2}+i\omega \gamma }
\end{equation}
($\gamma $ --- damping parameter) shifts the borders of the
disordered phase between AFM and stipe phases at $T=0$.

Nevertheless, our estimates and comparison of the available data
confirm, that the reasonable degree of tuning does
not lead to any topological modifications of the phase diagram.

\subsection{Summary}

To summarize, in the present work thermodynamic properties of of the 2D
$J_{1}-J_{2}$ $S=1/2$ Heisenberg model are considered for the entire phase
diagram the frames of one and the same approach --- spherically symmetric
self-consistent approach for two-time retarded Green's functions.

This work is supported by Russian Foundation for Basic Research, grant
13-02-00909a.

\end{document}